\documentclass[sigplan,10pt]{acmart}
\renewcommand\footnotetextcopyrightpermission[1]{}
\usepackage{fancyhdr}
\usepackage[normalem]{ulem}
\usepackage{microtype}
\usepackage{algorithm}
\usepackage{algpseudocode}
\usepackage[english]{babel}         
\usepackage[utf8]{inputenc}
\usepackage[T1]{fontenc}
\usepackage{colortbl}
\usepackage{multirow}
\usepackage{booktabs}
\usepackage{times}
\usepackage{tikz,pgfplots}
\usetikzlibrary{pgfplots.groupplots}
\usepackage{pgfplotstable}
\usepgfplotslibrary{external} 
\usetikzlibrary{patterns}
\usepgfplotslibrary{fillbetween}
\usetikzlibrary{fadings}
\usepackage{ifthen}

\AtBeginDocument{%
  \providecommand\BibTeX{{%
    \normalfont B\kern-0.5em{\scshape i\kern-0.25em b}\kern-0.8em\TeX}}}

\begin{document}

%
\title{Memory virtualization in virtualized systems: segmentation is better than paging}

%
\author{Boris TEABE }
\affiliation{Université de Toulouse}
\author{Peterson YUHALA}
\affiliation{
Université de Neuchâtel
}
\author{Alain TCHANA}
\affiliation{
ENS Lyon
}
%
\author{Fabien HERMENIER}
\affiliation{
Nutanix
}

\author{Daniel HAGIMONT}
\affiliation{
Université de Toulouse
}

\author{Gilles MULLER}
\affiliation{
Inria
}
              
%
%
%
%
%
%

%
\begin{abstract}
	The utilization of paging for virtual machine (VM) memory management is the root cause of memory virtualization overhead.
This paper shows that paging is not necessary in the hypervisor.
In fact, memory fragmentation, which explains paging utilization, is not an issue in virtualized datacenters thanks to VM memory demand patterns.
Our solution \textit{Compromis}, a novel Memory Management Unit, uses direct segment for VM memory management combined with paging for VM's processes.
The paper presents a systematic methodology for implementing \textit{Compromis} in the hardware, the hypervisor and the datacenter scheduler.
Evaluation results show that \textit{Compromis} outperforms the two popular memory virtualization solutions:
shadow paging and Extended Page Table by up to 30\% and 370\% respectively.
\end{abstract}

%
%

%
\settopmatter{printfolios=true}
\maketitle

\section{Introduction}
\label{introduction}
Virtualization has become the \textit{de facto} cloud computing standard because it brings several benefits such as optimal server utilization, security, fault tolerance and quick service deployment \citep{nhlearningsolutions,thrivenetworks,Armbrust:2010:VCC:1721654.1721672}.
However, there is still room for improvement, mainly at the memory level which represents up to 90\%~\cite{Yaniv} of the global
virtualization overhead.

Memory virtualization overhead comes from the necessity to manage three address spaces (application, guest OS and host OS) instead of two (application and OS) as in native systems.
Shadow paging \cite{Waldspurger} is the most popular memory virtualization solution.
Each page table inside the guest OS is shadowed by a corresponding page table inside the hypervisor, which contains the real mapping between Guest Virtual Addresses (GVA) and Host Physical Addresses (HPA).
Thus, shadow page tables are those used for address translation by the hardware page table walker (which resides inside the Memory Management Unit, MMU).
Page tables inside the guest OS are never used.

Shadow paging leads to one-dimensional (1D) page walk on TLB miss, as in a native system.
However, building shadow page tables comes with costly context switches between the guest OS and the hypervisor for synchronization.
Nested/Extended Page Table (EPT) \cite{Bhargava,Uhlig} has been introduced for avoiding page table synchronization cost.
It improves the page table walker to walk through two page tables (from the guest and from the hypervisor) at the same time in a 2D manner.
Thus, building the shadow page table does not require the protection of guest's page tables.
This drastically reduces the number of context switches.
However, this solution induces several memory accesses during address translation due to the 2D page walk mechanism.
In a radix-4 page table \cite{10.1007/978-3-540-39864-6_24} (the most popular case) for instance, this 2D page walk leads to 24 memory accesses on each TLB miss instead of 4 in a native system, resulting in significant performance degradation.


While many work proved the effectiveness of paging when dealing with processes (\emph{e.g.}, for reducing memory fragmentation), to our knowledge, there is no clear assessment of its effectiveness when dealing with virtual machines (VMs).
One explanation is that the implementation of hypervisors was inspired by the bare metal implementation of OSes.
By analyzing traces from two public clouds (Microsoft Azure and Bitbrain) and 308 private clouds (managed by Nutanix\footnote{Nutanix is a world wide private cloud provider.}), we show in Section \ref{analysis} that paging is not mandatory for managing memory allocated to VMs.
In fact, we found that memory fragmentation is not an issue in virtualized datacenters thanks to VM memory sizing and arrival rate.

This paper presents \textit{Compromis}, a novel MMU solution which uses segmentation for VMs and paging for processes inside VMs.
\textit{Compromis} allows a 1D page walk and generates zero context switch to virtualize memory.
\textit{Compromis} is inspired by \textit{Direct Segment} (DS) introduced by Basu \emph{et al.}~\cite{Basu} for native systems.
For memory hungry applications which allocate at start time their entire memory and self-manage it at runtime (\emph{e.g.}, Java Virtual Machine), their virtual memory space can be directly mapped to a large physical memory segment identified by a triple (\textit{Base}, \textit{Limit}, \textit{Offset}).
This way, the translation of a virtual address $va$ is given by a simple register to register addition (\textit{va+Offset}).
\textit{Compromis} generalizes DS to DS-n, allowing the provisioning of a VM with several memory segments.
In \textit{Compromis}, every processor context includes $2n$ new registers for address translation.
Contrary to other DS based solutions \cite{doItYourSelf,Gandhi2014}, \textit{Compromis} considers the entire VM memory and requires no guest OS and application modification.

This paper also investigates systems implications and presents a systematic methodology for adapting the hypervisor and other cloud services (\emph{e.g.}, datacenter scheduler) for making \textit{Compromis} effective.
To the best of our knowledge, this is the first DS based approach in virtualized systems which puts the entire puzzle together.

We have implemented a whole prototype in both Xen and KVM virtualized systems managed by OpenStack.
This means first the integration of our DS aware memory allocation algorithm.
Second the improvement of the Virtual Machine Control Structure (VMCS) data structure for configuring the new hardware registers introduced by DS-n.
Finally, the improvement of OpenStack's VM placement algorithm to minimize the number of memory segments.
For evaluations, since our solution relies on the modification of the hardware, we mimicked the functioning of a VM which runs on a DS-n machine as follows.
We run the VM in para-virtualized (PV) mode \cite{Barham03xenand} because the latter uses 1D page walk as DS-n.
However in PV, all page table modifications performed by the VM kernel trap into the hypervisor using hypercalls.
To avoid this behavior which will not exist on a DS-n machine, we have modified the guest kernel to directly set page table entries with the correct HPAs, calculated in the same way as a DS-n hardware would have done.
	
%

To be exhaustive, the paper makes the following contributions:
\begin{itemize}
	\item We first study the potential effectiveness of DS in virtualized datacenters.
	In order words, we answer this question: regarding VM memory demands, arrival times and departure times, is it likely to provision all or the majority of the VMs with one large memory segment?
	To answer this question, we study memory fragmentation in virtualized systems by analyzing traces from two public clouds (Microsoft Azure \cite{Cortez} and Bitbrain \cite{bitbrain}) and 308 private clouds.
	We found that using a DS aware memory allocation system, memory fragmentation is not a critical issue in virtualized datacenters as in native ones.	
	\item Drawing on this conviction, we propose DS-n, a generalization of DS to provision a VM with multiple memory segments.
	We present the necessary hardware modifications required by DS-n.
	\item We propose a DS aware VM memory allocation algorithm which minimizes the number of memory segments to use.
	\item We evaluated the performance gain of DS-n using an accurate methodology on a real environment.
	The main results are as follows.
First, the analyzed datacenter traces exhibit that it is then possible to provision up to 99.99\% of the VMs with one memory segment while three segments are sufficient to provision all the VMs.
Second, concerning the performance gain, DS-n reduces memory virtualization overhead to only 0.35\%, outperforming both shadow paging and EPT by up to 30\% and 370\% respectively.
The results also show that our memory allocation algorithm runs faster than traditional ones.
Xen's algorithm is outperformed by 80\%.
\end{itemize}

The remainder of the paper is as follows.
Section \ref{background} presents the necessary background to understand our contributions.
Section \ref{Assessment} evaluates the limitations of state-of-the-art solutions.
Section \ref{analysis} presents the analysis of several production datacenter traces and validate the opportunity to apply DS in virtualized systems.
Section \ref{contributions} presents the necessary hardware and software improvements to make DS-n effective.
Section \ref{evaluations} presents the evaluation results.
Section \ref{rw} presents the related work.
Section \ref{conclusion} concludes the paper.
\section{Background}
\label{background}
This section presents the two main techniques used to achieve memory virtualization, namely Shadow paging \cite{Waldspurger} and Extended Page Table (noted EPT) \cite{Bhargava,Uhlig}.

\subsection{Shadow paging}
\label{shadowPaging}
Shadow paging is a software memory virtualization technique.
In Shadow paging the hypervisor creates a \textit{shadow} Page Table (PT) for each guest PT.
This shadow PT holds the complete translation from GVA to HPA.
It is walked by the Hardware Page Table Walker (HPTW) on Translation Lookaside Buffer (TLB) miss.
Guest Page Tables (GPTs) are fake ones, they are not exploited.
To put in place shadow paging, the hypervisor write protects both the CR3 register (which holds the current PT address) and GPTs.
Each time the guest OS attempts to modify these structures, it then traps in the hypervisor which fix the CR3 register or the shadow PT.
Using shadow paging, the HPTW only performs a 1D page walk as in a native system, leading to 4 memory accesses.
However, the resulting context switches severely degrades the VM performance.

\subsection{Extended Page Table}
\label{ept}
EPT (also called Nested Page Table) is a hardware-assisted memory virtualization solution proposed by many chip vendors such as Intel and AMD.
It relies on a two layer PT.
The first PT layer resides in the guest address space and is exclusively managed by its OS, at the rate of one PT per process. This first layer PT thus contains GPAs which point to guest pages in the guest address space. Every process context switch triggers the setting by the guest OS of the CR3 register with the GPA of the scheduled-in process's PT address.
The second PT layer resides in the hypervisor, at the rate of one PT per VM. This PT represents the address space of the guest and it includes HPA which point to pages (real pages in RAM) in the host address space. Every vCPU context switch triggers the setting by the hypervisor of the nested CR3 register (nCR3) with the HPA of the scheduled-in VM's PT address.
On TLB miss, the hardware page table walker translates a virtual address \texttt{va} into the corresponding HPA by performing a 2-dimension page walk, leading to 24 memory accesses.

\section{Assessment}
\label{Assessment}
This section presents the overhead of memory virtualization in both native and virtualized systems.
Note that even in a native system, the expression "memory virtualization" is used because of the mapping of the process linear address space to the physical address space.

\begin{table}[!t]
\setlength{\tabcolsep}{3pt}
	\center
	\small
    \begin{tabular}{l|p{6cm}}
        \hline
        \textbf{Benchmark} & \textbf{Description}\\
        \hline
        \hline
        SpecCPU 2006 & Compute multi-threaded workloads\\        
        \hline
        PARSEC 3.0 & Compute multi-threaded workloads\\
        \hline        
        Redis & In-memory database\\
        \hline              
        Elastic search & In-memory database\\
        \hline
    \end{tabular}
    \caption{Benchmarks used for assessment and evaluation of our solution.}
    \label{tab:benchmarkList}
\end{table}

\begin{table}[!t]
\setlength{\tabcolsep}{3pt}
	\center
	\small
    \begin{tabular}{l|p{7cm}}        
        \hline
        \textbf{Native} & total cycles of all page walks\\
        \hline
        \textbf{EPT} & total cycles of all (GPT+EPT) walks\\
        \hline        
        \textbf{Shadow} & total cycles of all (hypervisor level PT walk+ VMEntry+VMExit+handler)\\
        \hline              
    \end{tabular}
    \caption{Formulas to estimate the overhead of memory virtualization. ("handler" is the handler which treats the VMExit generated when the guest OS attempts to modify the page table.)}
    \label{tab:assessmentMethodology}
\end{table}

\textbf{Methodology.}
Table \ref{tab:benchmarkList} lists the benchmarks used to evaluate the performance overhead of memory virtualization.
We do this while varying the memory page size in both the hypervisor and the guest OS.
This way, we also evaluated huge page-based solutions.
The notation \emph{\textbf{g}X}-\emph{\textbf{h}Y} means \emph{X} and \emph{Y} are respectively the memory page size in the \textbf{g}uest OS and the \textbf{h}ypervisor.
The evaluation metric is the time taken by both the hardware and the software for memory virtualization.
Table \ref{tab:assessmentMethodology} presents how the performance metric is calculated for each virtualization technology.
We rely on both Performance Monitoring Counters (PMC) and low-level software probes that we have written for the purpose of this paper.
Details on the experimental setup is given in Section \ref{methodologyperformanceGain}.

\begin{figure*}[ht!]
    \centering  
    \tiny 
\begin{tikzpicture}
  \begin{axis}[
 scale=0.9,
                height=5cm,
                width=21cm, 
    ybar ,   
	bar width=3pt,	
	enlargelimits=0.04,
	ylabel style={at={(0.04,0.5)}},
	ymin=0,	
	ymax=65,	
    legend style={at={(0.5,1.15)},
    legend cell align={left},
      anchor=north,legend columns=-1},
       nodes near coords,nodes near coords align={vertical},
every node near coord/.append style={color=black, rotate=90, anchor=west},
    ylabel={Virt. overhead (\%)},
    symbolic x coords={bzip2,gcc,mcf,gobmk,hmmer,sjeng,libquantum,h264ref,omnetpp,
    astar,xalancbmk,blackscholes,bodytrack,canneal,ferret,fluidanimate,streamcluster,vips,Redis,Elastic Search},
    xtick=data,
    x tick label style={rotate=90,anchor=east},
    ]
    \addplot +[draw=red, pattern color=red,pattern=dots,thick] coordinates {(bzip2,1.02138930319421) (gcc,2.50513014293662) (mcf,42.0725194269913) (gobmk,0.434411070922148) (hmmer,0.0259766073047252) (sjeng,2.05016856692594) (libquantum,9.84147953039356) (h264ref,0.437425170117029) (omnetpp,10.551767873157) (astar,8.31804307810766) (xalancbmk,9.15122504561278) (blackscholes,0.483208210084034) (bodytrack,0.649007908496732) (canneal,11.0637682659314) (ferret,5.31654590336134) (fluidanimate,5.49668550357829) (streamcluster,4.29735362815844) (vips,1.35722806862745) (Redis,6.46811178088235) (Elastic Search,7.39492416979988)};
    
    \addplot +[draw=red, pattern color=red,pattern=north east lines,thick] coordinates {(bzip2,13.0495954915969) (gcc,17.816091954023) (mcf,11.4298207335872) (gobmk,6.17283950617284) (hmmer,5.92417061611374) (sjeng,10.9157509230769) (libquantum,6.75891431116333) (h264ref,31.9672131147541) (omnetpp,6.51709401709401) (astar,29.5485636066723) (xalancbmk,29.8206277948229) (blackscholes,10.2546396201985) (bodytrack,35.1428144160311) (canneal,24.298264409685) (ferret,10.4761904761905) (fluidanimate,31.6058367331937) (streamcluster,45.2518046232059) (vips,48.868053967528) (Redis,42.8074651382239) (Elastic Search,50.933103137515)};				

    \addplot +[draw=blue, pattern color=blue,pattern=north west lines,thick] coordinates {(bzip2,5.14198004604758) (gcc,17.5326176713459) (mcf,23.648699585375) (gobmk,4.78359908883827) (hmmer,3.99032648125756) (sjeng,15.7927205428748) (libquantum,5.29729729729729) (h264ref,4.32276657060519) (omnetpp,7.89473684210526) (astar,10.2005231037489) (xalancbmk,8.07635829662262) (blackscholes,3.97155259998152) (bodytrack,4.63940193491645) (canneal,8.07618254670344) (ferret,3.2901436145932) (fluidanimate,4.45676730334771) (streamcluster,5.54828705722733) (vips,4.97237569060773) (Redis,34.6452604884283) (Elastic Search,19.3070028267757)};

     \addplot +[draw=blue, pattern color=blue, pattern=dots,thick] coordinates {(bzip2,5.0986030318) (gcc,15.7670002892) (mcf,18.9038381757) (gobmk,4.7684419308) (hmmer,3.9875144119) (sjeng,15.243491903) (libquantum,5.2447728912) (h264ref,4.2990788414) (omnetpp,7.1384117604) (astar,9.5323346442) (xalancbmk,7.8168107602) (blackscholes,3.9544853348) (bodytrack,2.7367592845) (canneal,7.5247333998) (ferret,3.1307123918) (fluidanimate,4.1116171853) (streamcluster,4.7832648812) (vips,4.9096532482) (Redis,31.3632583277) (Elastic Search,16.7937304089)};

  \legend{Native 4KB,SP g4KB,EPT g4KB-h2MB, EPT g2MB-h2MB}     
  \end{axis}  
\end{tikzpicture}
	\caption{Proportion of CPU time used for memory virtualization in native, virtualized shadow pagin (SP) and virtualized EPT.}
	\label{fig:assessment}
\end{figure*}
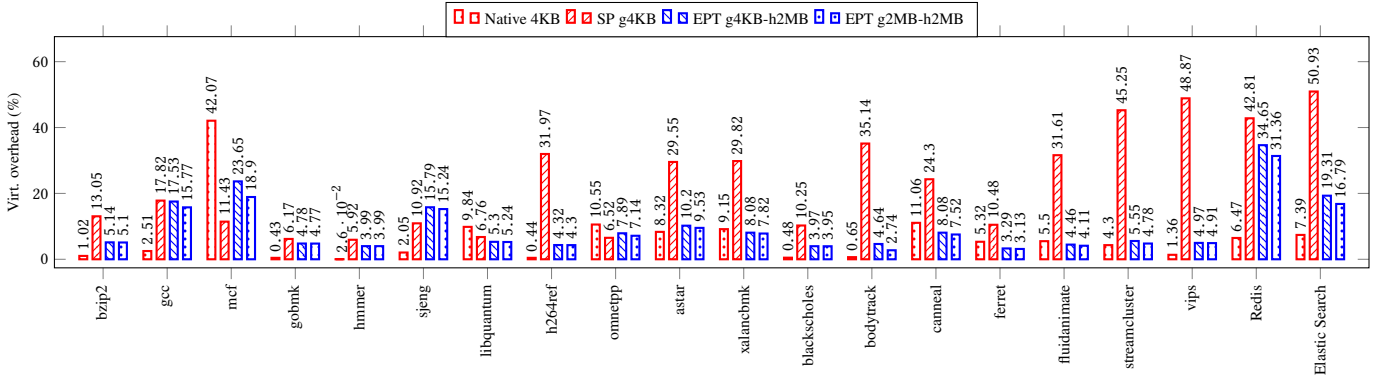

\textbf{Results.}
Figure \ref{fig:assessment} presents the results, interpreted as follows.
First, even in native systems memory virtualization takes a significant time proportion in the execution of an application, up to 42\% for mcf.
Second, running applications in a virtualized environment increases that duration, up to 50.93\% for Elastic Search under shadow paging.
Third, shadow paging incurs more overhead for the majority of applications than EPT, up to 43.89\% of difference for vips.
Finally, we can observe that even when huge pages are used simultaneously in the guest OS and the hypervisor, memory virtualization overhead is still high, almost 31.5\% for Redis benchmark.~\cite{doItYourSelf,Basu,Gandhi2014,Pham} have reached the same conclusion with the use of huge pages.

\textbf{Synthesis.}
These results show that the overhead of memory virtualization is very significant in a virtualized system even with huge pages.
The root cause of this overhead is the utilization of paging as the memory virtualization basis for VMs.


\section{Paging is not a fatality for VMs}
\label{analysis}
Several research work tried to reduce the overhead of memory virtualization in virtualized systems.
However, no work has questioned the relevance of paging in this context.
This section studies the (ir)relevance of paging when dealing with VMs.
To this end, we compare paging with segmentation, which is the alternative approach that has been left out.

Paging consists in organizing both the virtual address space of a process and the physical address space into fixed size memory chunks (4KB, 2MB, \emph{etc.}) called pages.
Thus, each virtual page can be housed in any physical page frame.
The process PT and the HPTW make it possible to find the actual mapping of a virtual page to a physical page address.
Segmentation on the other hand organizes both the virtual and the physical address spaces in the form of variable size memory chunks called segments.
The size of the latter is chosen by the programmer.
The correspondence between virtual segments and physical segments is provided by a segment table.
The virtual address to physical address mapping is done by a simple addition.

The main reasons which promote paging over segmentation in native systems are as follows:
($R_1$) Paging is invisible to the programmer; this is not the case with segmentation which hardens application programming.
($R_2$) Paging makes the implementation of the memory allocator in the OS easier.
Indeed, it only requires the use of a list of free pages and the choice of any page within this list upon receiving a memory allocation request is sufficient.
This is important for scalability.
With segmentation, there is a need to find the appropriate physical segment that satisfies the size of the virtual segment requested by the application.
($R_3$) Paging limits memory fragmentation \footnote {Internal fragmentation within pages is always possible, but it is negligible compared to the external fragmentation caused by segmentation}, which is not the case with segmentation.
($R_4$) Paging allows overcommitment, which is useful for optimizing memory utilization.

The question is whether all these reasons are valid when manipulating VMs.
To answer this question, we analyze the relevance of each reason in virtualized systems.
Before doing this, note that when we talk about memory management in a virtualized system, we are talking about the allocation of physical memory to VMs and not memory allocation to applications inside VMs.

\textbf{Relevance of $R_1$ (Segmentation hardens application programming)}.
This reason is valid in native environments (when dealing with applications) because application programmers do not have the expertise to manage segment size in a segmentation based system.
Moreover it is not the heart of the business logic of their application.
When dealing with VMs, developers are OS developers, who are expert.
Leaving OS developers the responsibility to manage memory segments is within their reach.

\textbf{Relevance of $R_2$ (Paging makes memory allocation easier)}.
It is necessary to facilitate the work of the memory allocator for scalability purposes.
In a native system, the memory allocator is subject to thousands of memory allocation and deallocation requests per second.
This is not true when dealing with VMs.
Each VM performs only one allocation (at startup) and deallocation (at shutdown).
Thus, the frequency of memory allocation and deallocation requests received by the hypervisor are not of the same order of magnitude as those received by the OS in a native system. Table \ref{tab:allocationNatifVirtuel} presents the average memory allocation frequency received of a server from a native system and virtualized private and public clouds (see Section \ref{evaluations} for more details on the analyzed datasets).
We observe a phenomenal difference between native and virtualized systems, which are quite stable.
Given the extremely low values for virtualized datacenters, the difficulty of finding free memory chunks does not mind with segmentation when dealing with VMs.
\begin{table}[!t]
\setlength{\tabcolsep}{3pt}
	\center
	\small	
    \begin{tabular}{l|l}
        \hline
        \textbf{Dataset} & \textbf{Alloc./Hour/Server} \\
        \hline
        \hline
        \textbf{Native - Our lab machine} & 82071.5 \\
        \hline        
        \textbf{Virtualized - Private clouds} & 0.056 \\  
        \hline              
        \textbf{Virtualized - Microsoft Azure public cloud} & 0.31 \\
        \hline
    \end{tabular}
    \caption{Memory allocation frequency (per hour on a server) in native and virtualized datacenters.}
    \label{tab:allocationNatifVirtuel}
\end{table}

\begin{figure*}[!h]
     	\centering 
     	\includegraphics[scale=.8]{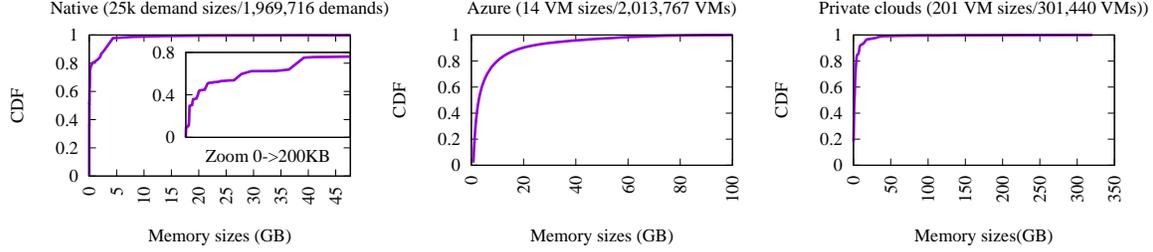}
     	\caption{CDF of Memory demand sizes in different datacenter types.}
     	\label{fig:tailleAllocation}
     \end{figure*}

\textbf{Relevance of $R_3$ (Paging limits fragmentation)}.
Fragmentation is due to the heterogeneity of memory demand sizes.
Indeed, a system in which all demand sizes are identical would not suffer from fragmentation.
To verify whether fragmentation could be a problem in a virtualized datacenter, we analyzed the memory demand sizes of the traces from the datacenters presented in Table \ref{tab:allocationNatifVirtuel}.
Figure \ref{fig:tailleAllocation} shows the CDFs.

We observe that public clouds stand out with very concentrated demand sizes (14).
This is because in public clouds VM sizes are imposed.
Things are slightly different in private clouds (201) where there is more freedom in VM size definition.
In contrary, demand sizes vary a lot in native systems (25k) than virtualized environments.
These results show that fragmentation is not a relevant issue when dealing with VMs.

\textbf{Relevance of $R_4$ (Paging allows memory overcommitment)}.
Overcommitment is a practice which allows to reserve more memory than the physical machine actually has.
It exploits the fact that all applications do not require their entire memory demand at the same time.
As a result, resource waste is avoided.
However, overcommitment comes with performance degradation (during memory reclamation) and performance unpredictability \cite{vladNitu}.
These limitations are acceptable in a native system because there is no contract between application owners and the datacenter owner; they both belong to the same company.
Best effort is the practice in such contexts.
Things are different in a virtualized datacenter, especially in commercial clouds.
In the latter, the datacenter operator should respect the contract signed with the VM owner, who paid for the reserved resources.
Therefore, even if a VM is not using its resources, these resources have already been amortized.
The necessity to avoid resource waste is less critical compared to a native system.
Futhermore, the implementation of overcommitment in a virtualized system is challenging because of the VM blackbox nature \cite{vladNitu}.
It requires an expertise in the workload and the system to configure it and react in case of performance issues.
As a consequence, no public clouds support it.
Private cloud providers either do not support it (Nutanix), disable it by default (VMWare TPS, Hyper-V dynamic memory) or enable it with extra warnings (RedHat with KVM).


\section{\textit{Compromis}: A DS based memory virtualization approach for VMs}
\label{contributions}
\textit{Compromis} is a hardware memory virtualization solution implemented within the MMU that exploits the strengths of both direct segment (DS) and paging.
The former is used by the hypervisor to deal with VMs while the latter is used by the guest OS to deal with processes.
The innovation is the utilization of DS instead of paging by the hypervisor.
Considering the fact that it may be impossible to satisfy a VM demand using a single memory segment, \textit{Compromis} generalizes DS to DS-n.
In the latter, a VM which is allocated \textit{k} segments, with $1 \leq k \leq n$, uses the \textit{Compromis} hardware feature.
This section presents the set of improvements that should be applied to the datacenter stack in order to make \textit{Compromis} effective.

\begin{figure}[!h]
    \centering 
    \includegraphics[scale=0.45]{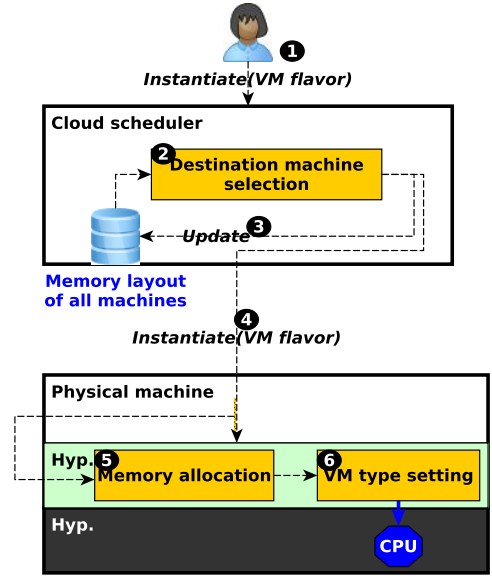}
	\caption{General functioning of a datacenter which implements \textit{Compromis}.}
	\label{fig:dsnGeneralOverview}
\end{figure}

\subsection{General overview}
\label{generalOverview}
Figure \ref{fig:dsnGeneralOverview} presents the general operations of a datacenter using \textit{Compromis}.
When a user requests a VM instantiation from a flavor (\#CPU, memory size), the cloud scheduler chooses the physical machine that will host the VM instance.
This choice is made according to a placement policy, which generally takes into account the resource availability constraints.
In a \textit{Compromis} aware datacenter, this policy is extended to choose the machine with the greatest chance of allocating large memory segments to the VM.
To this end, the cloud scheduler quickly simulates the execution of the memory allocator implemented by the hypervisor of compute nodes.
This simulation is built by the cloud scheduler on top of the current state of the memory layout of every machine that is locally stored and periodically updated (see Section \ref{cloudSchedulerLevelModification}).

When the hypervisor of the selected physical machine receives the VM instantiation request, it reserves the memory for the VM in the form of large memory segments rather than small page chunks as it is currently done (see Section \ref{hypervisorLevelModification}).
If the number of segments used to satisfy the VM is less than or equal to $n$, then the hypervisor configures the VM in DS-n mode, a new mode that the hardware implements (see Section \ref{hypervisorLevelImplementation}).
Otherwise, the VM is configured in shadow or EPT mode, depending on the datacenter operator.
In DS-n mode, the hardware performs an address translation by doing a 1D page walk (instead of 2D) followed by a series of register to register operations (see Section \ref{hardwareLevelModification}).
Notice that a \textit{Compromis} aware machine can simultaneously run DS-n and not DS-n VMs.
The next subsections detail the modifications that should be applied to each datacenter layer for building \textit{Compromis}.

\begin{figure*}[!h]
    \centering 
    \includegraphics[scale=0.3]{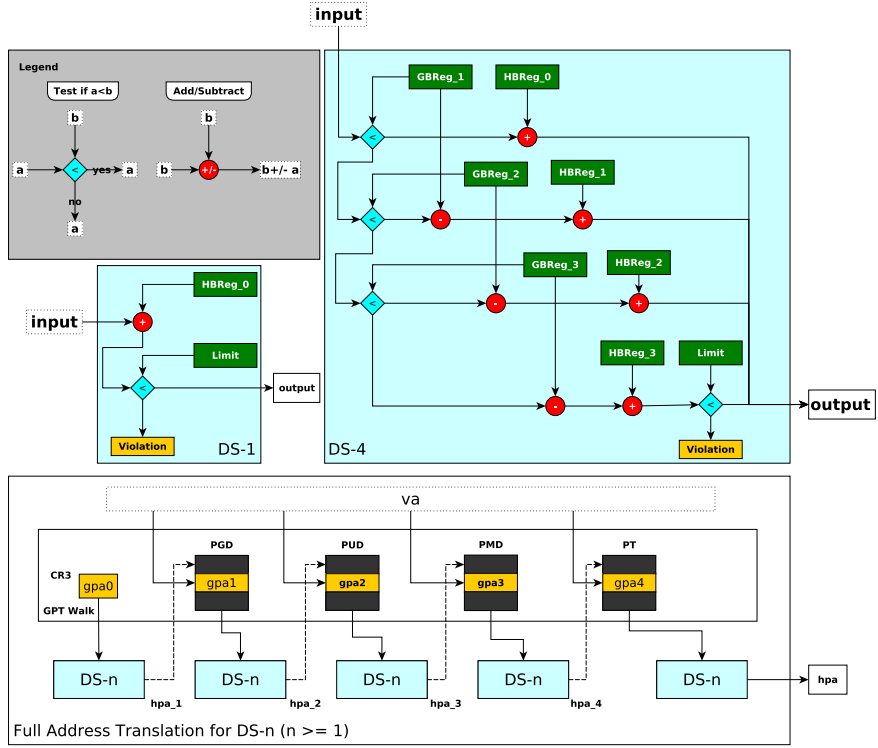}
	\caption{Address translation handling in two DS-n machine types (top DS-1 and bottom DS-4).}
	\label{fig:addressTranslationDSn}
\end{figure*}

\subsection{Hardware level contribution}
\label{hardwareLevelModification}
A hardware which implements \textit{Compromis} includes new registers to indicate the mapping of GPA segments (in the guest address space) to HPA segments (in the host address space).
The value of each register comes from a Virtual Machine Control Structure (VMCS), configured by the hypervisor at VM startup (see Section \ref{hypervisorLevelModification}).
The number of added registers is a function of $n$.
That is $n-1$ \textit{guest base registers} (noted $GBReg_1$, ..., $GBReg_{n-1}$, no such registers in DS-1), $n$ \textit{host base registers} (noted $HBReg_0$, ..., $HBReg_{n-1}$), and the \textit{limit} register.
These registers indicate the mapping as follows.
The GPA segment [$0, GBReg_1-1$] is mapped to the HPA segment [$HBReg_{0}, GBReg_1-1$].
The GPA segment [$GBReg_{i-1}, GBReg_i-1$] is mapped to the HPA segment [$HBReg_{i-1}, HBReg_{i-1}+(GBReg_i - GBReg_{i-1})$] (where\\
$GBReg_i - GBReg_{i-1}$ is the size of this segment).
For a VM with $k$ segments, the mapping of the last GPA segment
[$GBReg_{k-1}$, $GBReg_{k-1}$ $+ (limit - $ $HBReg_{k-1})$] is the HPA segment [$HBReg_{k-1},$ limit].
Once the configuration of the registers is made by the hypervisor, the translation of a virtual address \textit{va} to the corresponding HPA \textit{hpa} for a DS-n VM type (whose number of segment is lower than $n$) is summarized in Figure \ref{fig:addressTranslationDSn}.
Firstly, the MMU performs a 1D GPT walk, taking as input \textit{va}.
This operation returns a GPA \textit{gpa}.
Then \textit{hpa} is calculated as follows:
\begin{equation}
hpa=HBRreg_{i}+(gpa-GBRreg_{i})
\end{equation}
with [$GBRreg_{i}, *$] being the smallest GPA segment which contains \textit{gpa}.
If no such segment exists, a boundary violation is raised and trap in the hypervisor as a "DS-n violation" exception.
More generally, for each \textit{gpa} extracted from a GPT layer, an offset addition followed by a comparison is performed, meaning that every EPT walk is replaced by these two operations.
For instance, when the VM has only one segment, the computation of \textit{hpa} is as follows
\begin{equation}
hpa=HBRreg_{0}+gpa
\end{equation}
There is a boundary violation here if \textit{hpa} is greater than \textit{limit}.
The performance benefit of these operations against the 2D page walk done in EPT is discussed in Section \ref{evaluations}.

\subsection{Hypervisor level contribution}
\label{hypervisorLevelModification}
The hypervisor needs two main changes: the integration of a memory allocator for VMs and the configuration of the VMCS to indicate DS-n type VMs.

\subsubsection{A DS based memory allocator for VMs}
\label{dsBasedMemAllocator}
We assume that the physical memory is organized in two parts: 
the first part is reserved for hypervisor and privileged VM tasks while the second part is dedicated to user VMs.
This memory organization is found in almost all popular hypervisors.
In \textit{Compromis}, the first memory part is managed using the traditional memory allocator.
Concerning the second memory part, a new allocation algorithm is used to enforce large memory segment allocation to VMs.
This section describes this new allocator.

Implementing a memory allocator requires to answer three questions:
($Q_1$) which data structure to use for storing information about free memory segments?
($Q_2$) how do we choose elements from this data structure for responding to an allocation request?
($Q_3$) how do we insert an element into this data structure when there is a memory release?

\textbf{Answer to $Q_1$: data structure}.
We use a doubly linked list to describe free memory segments.
Each element in the list describes a segment using three variables:
\begin{itemize}
	\item \textit{base}: start address of the segment;
	\item \textit{limit}: end address of the segment;
	\item \textit{date}: allocation date of the segment containing \textit{base-1}.
\end{itemize}
The elements of the list are ordered in an ascending order of \textit{base}.
Hereafter, an element of the list is noted [$base$,$limit$,$date$].

\textbf{Answer to $Q_2$: allocation policy}.
When the hypervisor receives a request to start a VM with a memory demand $M$, it goes through the list described above to find out which segments should be allocated to the VM.
If it finds a segment of size $M$, then that segment is taken off the list and allocated to the VM.
Otherwise, the allocator chooses the largest segment $S_b$[$base$,$limit_i$,$date$] among segments which size is greater than $M$.
The VM is satisfied with a portion of $S_b$.
Note that taking the largest segment prevents the multiplication of small segments, which are bad for a DS based approach.
If there is no segment larger than $M$, then two options are possible.
The first option (\emph{Opt1}) satisfies the VM with the smaller segments.
This allows to give a chance to the VMs which will come later to have the big free segments.
The second option (\emph{Opt2}) chooses the largest segment that exists and executes the above algorithm with a new memory size $M'$, with $M'$ equals $M$ minus the size of the chosen segment.

\textit{Compromis} offers these two options to support various workload patterns and datacenter constitution.
A workload pattern is the set of VM instantiation and shutdown requests submitted to the datacenter during a period of time.
The constitution of a datacenter is the physical machine sizes.
The option selection is the responsibility of the cloud scheduler, which has a global view of the datacenter (see Section \ref{cloudSchedulerLevelModification}).

\textbf{Answer to $Q_3$: freed memory taken into account}.
Stopping a VM results in the free of its memory, which has to be inserted into the list of free memory segments.
Let $S$ be a memory segment to insert into the list.
The insertion is as follows.
If $S$ coincides with the beginning or the end of a segment $S'$ in the list, then $S'$ is simply extended (forward or backward).
If this extension causes the new big segment to coincide with the beginning or the end of the segments that follow or precede it, then the extension continues.
If there is no border coincidence between $S$ and the existing segments in the list, then $S$ is inserted in the list so that the ascending order is respected.


\subsubsection{VM type configuration}
\label{vmTypeConfiguration}
Let $k$ be the number of memory segments allocated to a VM.
If $k \leq n$ then the VM is of type DS-n, otherwise it is configured with EPT or shadow paging according to the datacenter administrator choice.
The type configuration of a VM is done by modifying the VMCS of its vCPUs.
To indicate that a VM is of type DS-n, a new bit of the \textit{Secondary Processor-Based VM-Execution Controls} is set.
Otherwise, this bit remains at zero.
For DS-n VMs, the hypervisor also positions new VMCS fields that will be used to populate \textit{GBReg}, \textit{HBReg}, and \textit{limit} registers.
The fields are populated in ascending order of crossing segments.
The value of the fields which map to \textit{HBReg} and \textit{limit} registers comes from the list of segments allocated to the VM.
Concerning the fields which map to \textit{GBReg} registers, their values are calculated as they are filled.
When $k < n$, the remaining fields are set to zero.

\subsection{Cloud scheduler level contribution}
\label{cloudSchedulerLevelModification}
The cloud scheduler is improved for two purposes:
a DS-n aware VM placement algorithm and memory allocation option selection.

\textbf{VM placement algorithm improvement.}
The placement algorithm determines the physical machine that will instantiate the VM.
Traditionally, it has an objective such as load balancing.
For example, the schedulers of OpenStack~\cite{filter-scheduler} and CloudStack~\cite{cloudstack} consist of a list of filters.
Each filter implements a concern such as resource-matchmaking~\cite{condor}, VM-VM or VM-host (anti-) affinities.
A filter receives as arguments a set of possible machine for the VM to boot and remove among them those not satisfying its concern.
Each time the scheduler is invoked to decide where to place a VM, it chains the filters to retrieve eventually the satisfying machines and pick one among them.

To take the benefit of DS-n, the VM scheduler must integrate inside its objective the maximization of the number of VMs of type DS-n.
For filter-based scheduler, it consists in implementing is a new filter to append to the existing list.
This filter maintains a local copy of the free memory segments on every machine and uses a simulator to evaluate the number of segments that will be used if the VM is instantiated on each machine.
It then selects the machine leading to the least number of segments.
For schedulers that do not rely on filters, the rational is to weight the existing objective against the one that consists in picking the machine minimizing the number of memory segments.
Note that this cloud scheduler modification does not affect (reduce) the hosting capacity of the datacenter because the destination machine is selected among the original cloud scheduler candidates.


\textbf{Memory allocation option selection.}
Section \ref{dsBasedMemAllocator} reported that the cloud scheduler has the responsibility to select the memory allocation option that all hypervisors will use.
To this end, it embeds a memory allocator simulator which implements the two options presented in \ref{dsBasedMemAllocator}.
Then it periodically (e.g., every week) replays in the simulator the recorded VM startup and shutdown logs.
This is done while varying the memory allocation option.
The selected option is the one that produces the large number of DS-n VMs.
All hypervisors are then notified with the name of the selected option and the logs repository is reset.

\subsection{Prototype}
\label{implementation}
We implemented \textit{Compromis} in two popular hypervisors (Xen and KVM), as well as in OpenStack's Nova scheduler.

\subsubsection{Implementation in the hypervisor}
\label{hypervisorLevelImplementation}
\textbf{Implementation in Xen.}
The implementation of \textit{Compromis} in Xen is straightforward.
First Xen already organizes the main memory in two parts as we wish.
The first part is managed by the Linux's memory allocator subsystem hosted within the privileged VM (\textit{dom0}).
The memory allocator for user VMs resides in the hypervisor core.
It is invoked by the \textit{dom0} during the VM instantiation process.
We simply replaced this allocator with the one described in Section \ref{dsBasedMemAllocator}.
We validated the effectiveness of this algorithm by starting VMs (in hardware-assisted virtualization (HVM) mode) with single segments, while the hypervisor still uses EPT for address translation.

Concerning the configuration of the VM type, the modification of Xen does not require any particular description other than what has been said in Section \ref{vmTypeConfiguration}.
Concerning the handling of cloud scheduler notifications related to the changing of the memory allocation option, we define a new hypercall that inform the hypervisor with the name of the selected option.

\textbf{Implementation in KVM.}
Unlike Xen, KVM does not hold memory in two blocks.
KVM relies on the Linux memory allocator which sees VMs as normal processes.
To implement \textit{Compromis} in KVM, we first enforce the organization of the physical memory in two blocks.
To this end, we use the cgroup mechanism.
Then the default Linux memory allocator is associated to the first block while our memory allocator manages the second block.
The \textit{/proc} file system is used to record the used memory allocation option imposed by the cloud scheduler.

\subsubsection{Implementation in the cloud scheduler}
\label{cloudSchedulerLevelImplementation}
The implementation of \textit{Compromis} in OpenStack Nova is quite straightforward because Nova's placement algorithm is easy to identify.
Its execution steps are also easy to identify, leading its extension with a simulation of our memory allocator very simple.
Concerning the periodical selection of the memory allocation option, we implemented a separate process which starts at the same time as Nova.
That process relies on existing OpenStack logs for obtaining VM startup and shutdown requests.

\subsection{Discussion}
\label{discussion}

\textbf{Memory overcommitment.}
Since \textit{Compromis} allows DS-n, the implementation of memory overcommitment is possible by performing dynamic segment resizing, addition, or removal, combined with a slight cooperation between the guest OS and the hypervisor.
A VM which needs more memory gains new segments or sees its segments extended.
Inversely, a VM which memory needs to be reduced will see either its segment sizes or number reduce.
The cooperation between the guest OS and the hypervisor is only necessary in this case.
In fact, the hypervisor should indicate to the guest OS the range of GPAs that should be released by the VM (using the balloon driver mechanism).
Indeed, the hypervisor is the only component which knows segment ranges.

\textbf{Memory Mapped IO (MMIO) region virtualization.}
IO device emulation and direct IO are the two IO virtualization solutions implemented by hypervisors in HVM mode.
The former solution, which is the most popular one, consists in protecting virtual MMIO ranges seen by the guest OS so that all IO operations performed by the guest trap in the hypervisor.
With this IO virtualization solution, the utilization of \textit{Compromis} is straightforward since virtual MMIO regions are at the GPA layer.
The validation step presented in Section \ref{hypervisorLevelImplementation} was performed under this solution.
With direct IO virtualization, the guest OS is directly presented the physical MMIO ranges configured by the hardware device.
This solution requires \textit{Compromis} to use several memory segments.
Note that this solution is not popular in todays clouds because it limits scalability (only enables very few virtual devices) and dynamic consolidation (VM live migration is not possible).

\section{Evaluations}
\label{evaluations}
We evaluated the following aspects:
(1) Effectiveness (see section \ref{dsIsPossible}): it is the capability to start a large number of VMs using the DS-n technology;
(2) Performance gain (see section \ref{performanceGain}): it is the capability to ameliorate the performance of applications which run in DS-n VMs;
(3) Startup impact (see Section \ref{overhead}): it is the potential positive/negative impact on VM startup latency.
Otherwise indicated, the used hypervisor and cloud management system are respectively Xen and OpenStack.

\subsection{Effectiveness}
\label{dsIsPossible}
Effectiveness evaluation is done by simulation using real datacenter traces.

\subsubsection{Methodology}
\label{dsIsPossiblemethodology}
We developed a simulator which mimics a datacenter managed with OpenStack \cite{filter-scheduler}, improved with our contributions.
The simulator replays VM startup and shutdown requests collected from several production datacenters, presented in Section \ref{dsIsPossibledataset}.
It considers that a VM demand includes a number of CPU cores and a memory size.
For each simulated VM startup request, the simulator logs two metrics:
the number of segments used for satisfying the VM memory demand and the time taken by our changes (extension of the cloud scheduler and the utilization of our memory allocation algorithm in the hypervisor).

To highlight the benefits of each \textit{Compromis} feature, we evaluated different versions including:
\begin{itemize}
	\item \textit{BaseLine}: the simulator implements both the native OpenStack scheduler and Xen's memory allocation algorithms;
	\item \textit{ImprovPlacement}: in this version the VM placement algorithm is improved to choose for every VM the machine which will use the minimum number of memory segments (as described in Section \ref{dsBasedMemAllocator});
	\item \textit{DynamicOptionSelec}: in this version the cloud scheduler calculates every week the best memory allocation option which will be used (as described in Section \ref{cloudSchedulerLevelModification}).
\end{itemize}

\subsubsection{Datasets}
\label{dsIsPossibledataset}
We used the traces of 2 public clouds (Bitbrains \cite{bitbrain} and Microsoft Azure \cite{Cortez}) and 308 private clouds.
Among other fields, each trace includes: the VM creation and destruction time, and the VM size (\#CPU and memory size).

\textbf{\textit{Bitbrains.}}
This cloud is a service provider specialized in managed hosting and business computation for many enterprises.
The dataset consists of 1,750 VMs, collected between August and September 2013.
%
Bitbrains does not include physical machine characteristics.


\textbf{\textit{Azure.}}
This is a public Microsoft cloud.
The dataset comprises $2,013,767$ VMs running on Azure from November 16$^{th}$, 2016 to February 16$^{th}$, 2017.

\textbf{\textit{Private clouds.}}
This group aggregates data of 308 private IaaS clouds running diverse workloads between November 1$^{st}$, 2018 to November 29$^{th}$, 2018.
For a given cloud, we collected one or more consistent snapshots of the cluster state at the moment the cluster triggered its hotspot mitigation service, which indicates that a machine is getting close to saturation.
A snapshot depicts the running VMs, their sizing (in terms of memory and cores) and their host (in terms of available memory and cores).
The collected dataset includes 301,440 VMs.
As the dataset contains snapshots and not the VM creation and destruction time, we derived from each snapshot a \emph{bootstorm} scenario where all the VMs are created simultaneously.
This dataset includes server characteristics.

\textbf{\textit{Composition and server characteristics used for Bitbrains and Azure.}}
Having no hardware information about the first two datasets, we consider that they are composed of server generations presented in Table \ref{tab:machineTypes}.
We chose these server generations as they are used in Azure according to this Youtube comment \cite{youtube}.
\emph{Gen6} and \emph{Godzilla} are new generations while \emph{Gen2} \emph{HPC}, \emph{Gen4} and \emph{Gen5} are older ones.
All server generations have the same proportion.

\begin{table}[!t]
\setlength{\tabcolsep}{3pt}
	\center
	\small	
    \begin{tabular}{l|l|l|l}
        \hline
        \textbf{Name} & RAM (GB) & Cores & \% in the traces\\
        \hline
        \hline
        \textbf{HPC} & 128 & 24 & 20\\
        \hline        
        \textbf{Gen4} & 192 & 24 & 20\\
        \hline              
        \textbf{Gen5} & 256 & 40 & 20\\
        \hline              
        \textbf{Gen6} & 192 & 48 & 20\\
        \hline              
        \textbf{Godzilla} & 512 & 32 & 20\\
        \hline                      
    \end{tabular}
    \caption{Server generations used in the replay of Bitbrains and Azure traces.}
    \label{tab:machineTypes}
\end{table}

\subsubsection{Results}
\label{dsIsPossibleresults}
\textbf{Bitbrain and Azure - Table \ref{tab:dsIsPossibleresults}}.
\textit{BaseLine} provides better results in Bitbrain (up to 81\% of VMs are satisfied with less than four memory segments) comparing to Azure (only about 24\% of VMs are satisfied with less than four memory segments).
This is because the VMs running on Bitbrain have a longer life time than Azure.
However, our solutions satisfy more VMs than \textit{BaseLine} (99.95\%-100\%).
This is because \textit{BaseLine}, which implements Xen, organizes the physical memory in the forms of small memory chunks which are then used for allocation.
As a naive algorithm, Xen cannot enforce DS to a VM even if it exists a free memory segment which is larger than the memory demand.
In contrast, \textit{Compromis} enforces DS near to the perfection (more than 99\% of VMs are satisfied with only one memory segment).
Our two memory application options discussed in \ref{dsBasedMemAllocator} show their slight difference in Bitbrain: \textit{ImprovPlacement+Opt1} satisfies more VMs with only one segment in comparison with \textit{ImprovPlacement+Opt2}.
Finally, dynamically switching between the two options (\textit{DynamicOptionSelec}) is the best solution (99.99\% of VMs use one memory segment).

\begin{figure}[!h]
\centering 
     	\includegraphics[scale=.7]{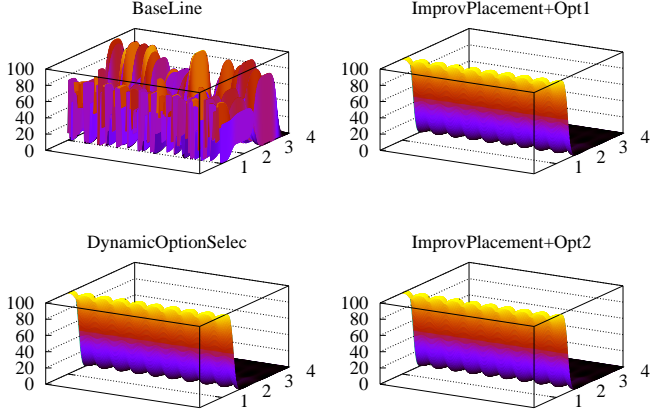}
     	\caption{Number of memory segments allocated to VMs from 308 private clouds (longitude=cloud, latitude=\#segment, depth=proportion).}
     	\label{fig:dsIsPossibleresultsNutanix}
     \end{figure}

\textbf{Private clouds - Figure \ref{fig:dsIsPossibleresultsNutanix}}.
We plot the results for these clouds separately from the previous ones because of the multitude number of clouds.
We can make the same observation as above.
Our solutions satisfy almost all VMs with only one memory segment, see a kind of wall at 1 on the latitude axis.
\begin{table}[!t]
\setlength{\tabcolsep}{3pt}
	\center
	\small	
    \begin{tabular}{l|l|l|l|l}
        \hline
        \multicolumn{5}{|c|}{\textbf{Bitbrain}}\\ 
        \hline
        \hline  
        \textbf{Solution} & \textbf{1 seg.} & \textbf{2 seg.} & \textbf{3 seg.} & \textbf{>3 seg.}\\             
        \hline          
        \textbf{BaseLine} & 12.816 & 44.376 & 30.078 & 12.728\\
        \hline        
        \textbf{ImprovPlacement+Opt1} & 100 & 0 & 0 & 0\\
        \hline              
        \textbf{ImprovPlacement+Opt2} & 100 & 0 & 0 & 0\\
        \hline              
        \textbf{DynamicOptionSelec} & 100 & 0 & 0 & 0\\
        \hline
    \end{tabular}
    
    \begin{tabular}{l|l|l|l|l}
        \hline
        \multicolumn{5}{|c|}{\textbf{Azure}}\\ 
        \hline
        \hline  
        \textbf{Solution} & \textbf{1 seg.} & \textbf{2 seg.} & \textbf{3 seg.} & \textbf{>3 seg.}\\             
        \hline          
        \textbf{BaseLine} & 3.581 & 11.171 & 9.996 & 75.252\\
        \hline        
        \textbf{ImprovPlacement+Opt1} & 99.9736 & 0.026 & 0 & 6.18E-05\\
        \hline              
        \textbf{ImprovPlacement+Opt2} & 99.947 & 0.007 & 0.022 & 0.021\\
        \hline              
        \textbf{DynamicOptionSelec} & 99.999 & 7.07E-04 & 0 & 0\\
        \hline
    \end{tabular}    
     	\caption{Number of memory segments allocated to VMs from Bitbrain and Azure.}
     	\label{tab:dsIsPossibleresults}
\end{table}

\subsection{Performance gain}
\label{performanceGain}
This section evaluates the performance gain brought by the utilization of DS-n.

\subsubsection{Methodology}
\label{methodologyperformanceGain}

A DS-n machine handles a TLB miss using a 1D page walk follows by a set of register to register operations.
We mimic this functioning by running the VM in para-virtualized (noted PV) mode \cite{Barham03xenand} which also uses a 1D page walk.
However in PV, all page table modifications performed by the VM kernel trap into the hypervisor using hypercalls.
We modified the guest kernel to directly set page table entries with the correct HPAs, calculated in the same way as a DS-n hardware would have done.
The reader could legitimately ask why using PV to simulate a hardware-assisted solution.
We claim that our approach makes sense in our context because the benchmarks do not solicit PV machinery: all disks are in-memory (\textit{tmpfs}) based and all network requests use the \textit{loopback} interface.
Accordingly, only the memory subsystem is solicited.

The evaluation methodology we use is as follows.
Let $T_{1D}$ be the execution time of the VM in this modified PV context.
We estimate the cost (noted $T^n_{reg2reg}$) of the register to register operations performed by the DS-n hardware on TLB miss using an assembly code which executes that operations.
It is adaptable according to the value of $n$.
Let $N_{tlb}$ be the number of TLB misses (collected using PMC) generated by the application when it is executed in a native system.
We estimate the execution time $T_{DS-n}$ of a VM on a DS-n using this formula
\begin{equation}
	T_{DS-n}=T_{1D} + N_{tlb} \times T^n_{reg2reg}
\end{equation}
We evaluated different values of $n$ from 1 to 3.
We compare DS-n with EPT (in which the execution time is noted $T_{ept}$) and shadow paging (in which the execution time is noted $T_{sha}$). We used 4KB page size in guest VMs as is the standard size. The characteristic of the experimental machine is presented in Table \ref{tab:machineCharacteristics}.
Notice that this machine includes a page walk cache \cite{Barr}.
The list of benchmarks we use (as previous work) are presented in Table \ref{tab:benchmarkList}.
Each benchmark runs in a VM having a single vCPU and 5 GB memory.
The used hypervisor and OS are Xen 4.8 and Ubuntu 16.04 (Linux kernel 4.15) respectively.
\begin{table}[!t]
\setlength{\tabcolsep}{3pt}
	\center
	\small
    \begin{tabular}{l|l}
        \hline
        Processor & Single socket Intel(R) core (TM) i7-3768\\
        		& @2.40GHz 4cores\\
        \hline
        Memory & 16GB DDR4 1600MHz\\
        \hline                
        DTLB & 4-way, 64 entries\\
        \hline              
        ITLB & 4-way, 128 entries\\
        \hline                      
    \end{tabular}
    \caption{Characteristics of the experimental machine.}
    \label{tab:machineCharacteristics}
\end{table}

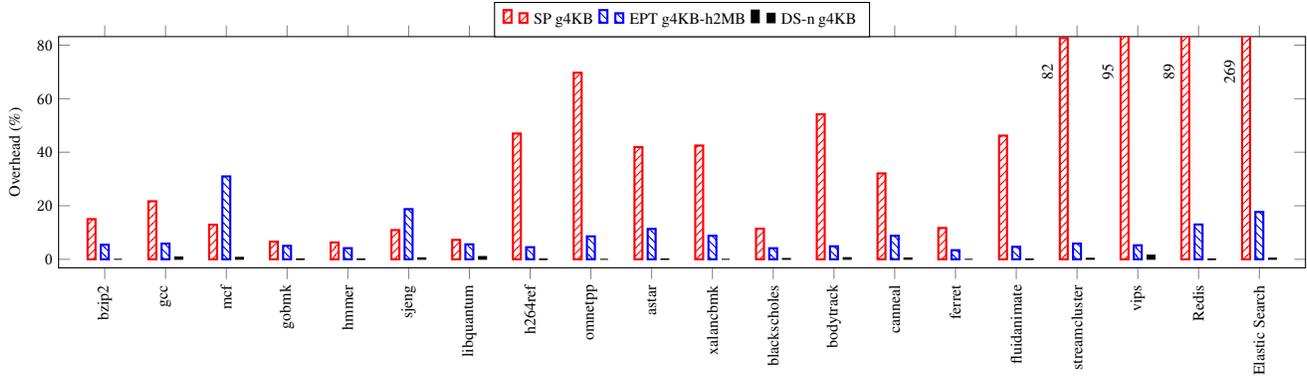
\begin{figure*}[ht!]
    \centering  
    \tiny   
\begin{tikzpicture}
  \begin{axis}[
 scale=0.9,
                height=5cm,
                width=20cm, 
    ybar ,   
	bar width=3pt,	
	enlargelimits=0.04,
	ylabel style={at={(0.04,0.5)}},
	ymin=0,	
	ymax=80,	
    legend style={at={(0.5,1.15)},
    legend cell align={left},
      anchor=north,legend columns=-1}, 
    ylabel={Overhead (\%)},
    symbolic x coords={bzip2,gcc,mcf,gobmk,hmmer,sjeng,libquantum,h264ref,omnetpp,
    astar,xalancbmk,blackscholes,bodytrack,canneal,ferret,fluidanimate,streamcluster,vips,Redis,Elastic Search},
    xtick=data,
    x tick label style={rotate=90,anchor=east},
    ]
  
    \addplot +[draw=red, pattern color=red,pattern=north east lines,thick] coordinates {(bzip2,15.008090606796) (gcc,21.678321678322) (mcf,12.904818335308) (gobmk,6.578947368421) (hmmer,6.297229219144) (sjeng,10.915750923077) (libquantum,7.248858441781) (h264ref,46.987951807229) (omnetpp,69.71428571429) (astar,41.941747563107) (xalancbmk,42.492012763578) (blackscholes,11.426372992209) (bodytrack,54.184920451925) (canneal,32.097367676196) (ferret,11.702127659574) (fluidanimate,46.211306964746) (streamcluster,82.654422327109) (vips,95.572450805009) (Redis,89.850360591096) (Elastic Search,269.051409989514)};				

    \addplot +[draw=blue, pattern color=blue,pattern=north west lines,thick] coordinates {(bzip2,5.42071197411) (gcc,5.856643356643) (mcf,30.973538704581) (gobmk,5.023923444976) (hmmer,4.156171284635) (sjeng,18.754578754579) (libquantum,5.593607305936) (h264ref,4.518072289157) (omnetpp,8.571428571429) (astar,11.359223300971) (xalancbmk,8.785942492013) (blackscholes,4.135808406271) (bodytrack,4.865114134194) (canneal,8.785734503255) (ferret,3.402076828117) (fluidanimate,4.664660361135) (streamcluster,5.87420480197) (vips,5.232558139535) (Redis,13.007097209253) (Elastic Search,17.682041101398)};

    \addplot +[draw=black, pattern color=red,fill=black] coordinates {(bzip2,0) (gcc,0.835664335664) (mcf,0.710900473934) (gobmk,0.05980861244) (hmmer,00.062972292191) (sjeng,0.527472527473) (libquantum,1.027397260274) (h264ref,0.033467202142) (omnetpp,0) (astar,0.097087378641) (xalancbmk,0) (blackscholes,0.288544772531) (bodytrack,0.611021443394) (canneal,0.481177469573) (ferret,0) (fluidanimate,0.049441100602) (streamcluster,0.364252000821) (vips,1.53335037056) (Redis,0.0123) (Elastic Search,0.4)};
  \legend{SP g4KB,EPT g4KB-h2MB, DS-n g4KB}     
\node[above,black,rotate=90] at (157,70) {82};  
\node[above,black,rotate=90] at (167,70) {95};  
\node[above,black,rotate=90] at (177,70) {89};  
\node[above,black,rotate=90] at (187,70) {269};  
  \end{axis}  
  
\end{tikzpicture}
	\caption{Performance overhead of DS-n compared with shadow paging (SP) and EPT. Lower is better.}
	\label{fig:resultsperformanceGain}
\end{figure*}

\subsubsection{Results}
\label{resultsperformanceGain}
Figure \ref{fig:resultsperformanceGain} presents the evaluation results.
We only present the results for DS-1 because we obtained almost the same results with DS-2 and DS-3.
This is because the cost of register to register operations realized in DS-1, DS-2 and DS-3 is extremely low compared with the cost of a 2D page walk.

Figure \ref{fig:resultsperformanceGain} is interpreted as follows.
First, obviously CPU intensive only applications (\emph{e.g.}, hmmer from PARSEC) do not benefit enough from DS-n.
Second, we confirm that DS-n almost nullifies the overhead of memory virtualization and leads the application almost to the same performance as in native systems.
In fact, all black histogram bars are very close to 1.
DS-n outperforms both EPT (up to 30\% of performance difference for mcf) and shadow paging (up to 370\% of performance difference for Elastic Search).
Finally, we observe that DS-n produces a very low, close to zero, overhead (0.35\%) but also a stable overhead (0.42 standard deviation).
While a smaller overhead is always appreciable, a stable overhead can also be a requirement to host latency sensitive applications like databases or real-time systems.

To justify the origin of this significant performance gap between these memory virtualization technologies, we analyzed the values of the internal metrics focusing on applications Redis, gcc, and Elastic Search.
For DS-n, the cost of memory virtualization is $C_{DS-n}=C_{1D} \times N^{DS-n}_{tlb}$, where
$C_{1D}$ is the number of CPU cycles for performing a 1D page walk and $N^{DS-n}_{tlb}$ is the number of TLB misses.
For EPT, that cost is $C_{EPT}=C_{2D} \times N^{EPT}_{tlb}$,
where $C_{2D}$ is the number of CPU cycles used to perform a 2D page walk and
$N^{EPT}_{tlb}$ is the number of TLB misses.
For shadow paging, the cost is $C_{Sha}=C_{1D} \times N^{Sha}_{tlb} + N^{Sha}_{exit} \times (C_{exit} + C_{enter} + C_{handler})$, where
$N^{Sha}_{tlb}$ is the number of TLB misses;
$N^{Sha}_{exit}$ is the number of VMExit related to page table modification operations;
$C_{exit}$ is the cost for performing VMExit followed by VMEnter;
and $C_{handler}$ is the average execution time of memory management handlers in the hypervisor.
Table \ref{tab:resultsDetails} presents the values of these costs, according to our experimental machine.
We observe that $C_{DS-n}$ is very lower than $C_{EPT}$ (e.g., $\times 6$ for Redis) and $C_{Sha}$ ($\times 14$).
\begin{table}[!t]
\setlength{\tabcolsep}{3pt}
	\center
	\small
    \begin{tabular}{l||l|l|l}
        \hline
        \textbf{Technology} & \textbf{Redis} & \textbf{gcc} & \textbf{Elastic Search}\\
        \hline
        \hline
         $C_{DS-n}$ & 3 & 13 & 14\\        
        \hline
        $C_{EPT}$ & 17 & 17 & 46\\
        \hline        
        $C_{Sha}$ & 25 & 62 & 201\\
        \hline
    \end{tabular}
    \caption{The total cost (in second) of each memory virtualization technology for Redis, gcc and Elastic Search.}
    \label{tab:resultsDetails}
\end{table}

\begin{table}[!t]
\setlength{\tabcolsep}{3pt}
	\center
	\small
    \begin{tabular}{l||l|l|l}
        \hline
        \textbf{Solution} & \textbf{Bitbrain} & \textbf{Azure} & \textbf{Private clouds}\\
        \hline
        \hline
        \textbf{BaseLine} & 6.42-139.47 & 17.76-520.55 & 1.92-18.27\\
        \hline        
        \textbf{DynamicOptionSelec} & 3.57-1.23 & 3.42-1.18 & 0.098-0.011\\
        \hline
    \end{tabular}
     	\caption{Memory allocation latency (mean-stdev) in ms.}
    \label{tab:overhead}
\end{table}
\subsection{Startup impact}
\label{overhead}
Recall that \textit{Compromis} extends the Cloud scheduler (which intervens on VM startup time) and changes the default memory allocator used by the hypervisor (also at VM start time).
Therefore, one may legitimately ask where these changes impact the VM startup latency.
We answer this question by summing the cost of the extension with the cost of our memory allocation algorithm, the we compare it with the cost of the default Xen memory allocation algorithm.
We rely on simulation logs generated during the evaluations presented in Section \ref{dsIsPossible}.
The experiment reports that almost all the different versions of our solution have the same complexity, thus we only present the results for \textit{DynamicOptionSelec} in Table \ref{tab:overhead}.
These results are interpreted as follows.
First, we observe that our solution reduces the startup time, by up to 80\% for Azure VMs.
The is because our allocation algorithm is simpler with regards to Xen which organizes memory in several memory chunk lists and iterate over these lists several times to satisfy memory demand.
Second, the smaller standard deviation reports that the startup time becomes more stable than Xen.
Such a predictability is critical for auto-scaling services, as demonstrated by Nitu \emph{et. al.}~\cite{Nitu2017SQB}.
The unpredictability of Xen comes from its complex memory allocation algorithm presented above.


\section{Related work}
\label{rw}
The overhead of memory virtualization in native systems has been proven by several previous work \cite{Basu,Basu2012,Bhattacharjee,Yaniv,Barr,Bhattacharjee2013,Park2017,Marathe,Cox2017,Bhattacharjee2017TP,Panwar,Haria}.
It has also been shown that this overhead is exacerbated in virtualized environments \cite{Gandhi,Bhargava,Uhlig,Yaniv,Ahn,Adams,doItYourSelf,Gandhi,Agesen,Wang2011,Gandhi2014,Chang2013,Pham}.
This section presents existing work in the latter context.
The research in this domain can be classified into two categories:
software and hardware-assisted solutions.

\subsection{Software solutions}
\label{sbs}
Direct paging \cite{directPaging} is similar to shadow paging \cite{Waldspurger} (presented in Section \ref{shadowPaging}), but it requires the modification of the guest OS.
In Direct paging \cite{directPaging}, the hypervisor introduces an additional level of abstraction between what the guest sees as physical memory and the underlying machine memory.
This is done through the introduction of a Physical to Machine (P2M) mapping maintained within the hypervisor such as in shadow paging.
The guest OS is aware of the P2M mapping and is modified such that instead of writing PTE it would instead write entries mapping virtual addresses directly to the machine address space by using itself the P2M.
As shadow paging, direct paging uses a 1D page walk to handle a TLB miss.
However, it includes two main drawbacks:
context switches between the guest and the hypervisor for building the P2M table, and
the modification of the guest OS (making proprietary OSes such as Windows not usable).

\subsection{Hardware-assisted solutions}
\label{hbs}
Both Intel and AMD proposed EPT \cite{Bhargava,Uhlig}, a hardware-assisted solution which does not include software solution's limitations.
We have already presented this solution in Section \ref{ept}.
As shown in the latter, EPT is far from satisfactory because of the 2D page walk that it imposes.
To reduce the overhead caused by this 2D page walk, several works have proposed the extension of the page walk cache (PWC) \cite{Barr}, used in native systems.
Such a cache avoids page walk on PWC hit.
Bhargava et al. \cite{Bhargava} investigated for the first time this extension of PWC for EPT.
The main limitation of such solutions is their inefficiency facing large working set size VMs (e.g., in-memory databases) \cite{Yaniv}.
Also, PWC based solutions suffer from a high rate of cache misses when several VMs share the same machine due to cache evictions.
Ahn et al. \cite{Ahn} used a flat EPT instead of the traditional multi-level radix.
By this way, the authors reduced the number of memory references on TLB miss to 9.
\textit{Compromis} totally eliminates the EPT, resulting in 4 memory references for each TLB miss.

Some solutions improved the TLB \cite{Pham,Yaniv,Ryoo}.
Ryoo et al. \cite{Ryoo} presented \textit{POM-TLB}, a very large level-3 in RAM TLB.
\textit{POM-TLB} brings two main advantages.
First, the number of TLB misses is reduced because of the large TLB size, thus reducing the number of 2D page walks.
Second, \textit{POM-TLB} benefits the data cache to reduce RAM references.
However, on cache miss a RAM access is necessary.
Also, on \textit{POM-TLB} miss, the hardware is still performed a 2D page walk.
This solution can be used at the same time with \textit{Compromis}.

Wang et al. \cite{Wang2011} and Gandhi et al. \cite{Gandhi} showed that neither EPT nor shadow paging can be a definite winner. 
They proposed dynamic switching mechanisms that exceed the benefits of each technique.
To this end, TLB misses and guest page faults are monitored to determine the best technique to apply.
Such dynamic solutions come with a significant overhead related to two tasks:
the monitoring and the computation of considered metrics consume a lot of CPU cycles, and
switching from one technique or another requires to rebuild new page tables.

\subsection{Orthogonal solutions}
\label{orthogonal}
Some researchers like Kwon et al. \cite{Kwon,Kwon2016} proposed the utilization of huge pages \cite{Shanley,thp} in the guest OS and the hypervisor at the same time.
This way, the number of hierarchy in the page table is reduced, thus the number of memory references during page walk is reduced too.
However, using huge pages leads to two main limitations for the guest.
First, it increases memory fragmentation, thus memory waste for the guest.
This could lead to a memory pressure in the guest OS, resulting in swapping, which is negative for application performance.
Second, huge pages increase average and tail memory allocation latency in the guest because zeroing a huge page at page allocation time is more time consuming than zeroing a 4Kb page. 

Talluri et al. \cite{Talluri} proposed \textit{Hashed page tables} in native systems as an efficient alternative to the radix page table structure.
With hashed page tables, address translation is done using a single memory reference, assuming no collision.
Yaniv et al. \cite{Yaniv} presented how this technique can be adapted for virtualized systems.
The authors showed that by using a 2D hashed page table hierarchy, the page walk is done with 3 memory references instead of 24.
This is one less than in \textit{Compromis} and native systems but suffers from hash collisions.

\textbf{Direct segment (DS) based solutions.} Previous work showed the benefits of DS in both native \cite{Kunati,Basu} and virtualized systems \cite{doItYourSelf,Gandhi,Gandhi2014}.
Alam et al. presented DVMT\cite{doItYourSelf}, a mechanism which allows applications inside the VM to request DS allocations directly from the hypervisor.
The application is responsible for mapping GVAs which are in the allocated DS address space.
This is a limitation for application developers who are not expert.
Ganghi et al. \cite{Gandhi2014} proposed three memory virtualization solutions based on DS.
Their \textit{VMM Direct} mode is very close to \textit{Compromis}, but DS does not concern the entire VM memory.
In addition, the authors mainly investigated the two other modes.

More generally, existing solutions in this category mainly focused on hardware contributions while we study the entire cloud stack consequences.
Also, they relied only on simulations while we tried to perform accurate experiments on real machines using real systems.
Finally, we motivate (relying on trace analysis) for the first time the relevance of DS for VMs.
\section{Conclusion}
\label{conclusion}
This paper presented \textit{Compromis}, a novel MMU solution for virtualized systems.
\textit{Compromis} generalizes DS to provide the entire VM memory space using a minimal number of memory segments.
This way, the hardware page table walker performs a 1D page walk as in native systems.
By analyzing several production datacenter traces, the paper showed that \textit{Compromis} provisioned up to 99.99\% VMs with a single memory segment.
The paper presented a systematic implementation of \textit{Compromis} in the hardware, the hypervisor and the cloud scheduler.
The evaluation results show that \textit{Compromis} reduces the memory virtualization overhead to only 0.35\%.
Furthermore, \textit{Compromis} reduces the VM startup latency by up to 80\% while providing also a predictable value.

\bibliographystyle{ACM-Reference-Format}
\bibliography{main}

\end{document}